# Bias field effect on the temperature anomalies of dielectric permittivity in PbMg$_{1/3}$Nb$_{2/3}$O$_3$- PbTiO$_3$ single crystals

I. P. Raevski[1], S. A. Prosandeev[1], A. S. Emelyanov[1], S.I. Raevskaya[1], Eugene V. Colla[2], D. Viehland[3], W. Kleemann[4], S. B. Vakhrushev[5], J-L. Dellis[6], M. El Marssi[6], L. Jastrabik[7]

[1]Physics Department and Research Institute of Physics, Rostov State University, 344090 Rostov on Don, Russia
[2]Department of Physics, University of Illinois at Urbana Champaign, Urbana, Illinois 61801
[3]Department of Materials Science and Engineering, Virginia Tech, Blacksburg, Virginia 24061
[4]Angewandte Physik, Universität Duisburg-Essen, D-47048 Duisburg, Germany
[5]A. F. Ioffe Physical-Technical Institute,194021 St. Petersburg, Russia
[6]LPMC, Universite de Picardie Jules Verne, 33, rue Saint Leu, F-80039 Amiens, France
[7]Institute of Physics AS CR, Na Slovance 2, 182 21 Prague 8, Czech Republic

In contrast to ordinary ferroelectrics where the temperature, $T_m$, of the permittivity maximum monotonically increases with bias field, $E$, in (1-$x$)PbMg$_{1/3}$Nb$_{2/3}$O$_3$-($x$)PbTiO$_3$ (0≤$x$≤0.35) single crystals, $T_m$ was found to remain constant or to decrease with $E$ up to a certain threshold field, $E_t$, above which $T_m$ starts increasing. $E_t$ decreases with $x$ and almost disappears at about $x$ = 0.4. We explain this dependence in the framework of models, which take into account quenched random fields and random bonds. For crystals with 0.06≤$x$≤0.13, the T-E phase diagrams are constructed. In contrast to PMN, they exhibit an additional, nearly field-independent boundary, in the vicinity of the Vogel-Fulcher temperature. We believe this boundary to correspond to an additional phase transition and the appearing order parameter is likely to be nonpolar.

## I. INTRODUCTION

Solid solutions of (1-$x$)PbMg$_{1/3}$Nb$_{2/3}$O$_3$–($x$)PbTiO$_3$ (PMN-$x$PT) have been in the focus of materials research for almost two decades due to their superior dielectric, electrostrictive and piezoelectric properties [1-9] . Besides, they serve as model objects for studying relaxor properties. Relaxors represent a class of disordered crystals exhibiting a broad and frequency-dependent temperature maximum of the dielectric permittivity vs. temperature instead of a sharp maximum inherent to normal ferroelectrics.

Ferroelectric properties are expected when a relaxor is subjected to an electric field above a boundary in the T-E phase diagram, separating the relaxor and ferroelectric phases. T-E-phase diagrams have been constructed in several studies for [111]-oriented PbMg$_{2/3}$Nb$_{1/3}$O$_3$ (PMN) crystals [4,10,11]. The main conclusion stemming from experiment [4] is that the phase transition between the relaxor and ferroelectric phases in PMN is first order. This phase transition was modeled by Vugmeister and Rabitz [12] by considering the average polarization as a function of field in the spirit of the Landau mean-field theory. Pirc, Blinc, and Kutnjak [13] modeled a similar phase diagram obtained for Pb$_{1-y}$La$_y$Zr$_{1-x}$Ti$_x$O$_3$ (PLZT) within the Spherical Random Bond Random Field (SRBRF) model supplemented with a field modulation of intercluster coupling. The present study will present experimental phase diagrams for several compositions of PMN-$x$PT in the range 0.13>$x$>0 expanding the conclusion about the first order phase transition under bias field.

It is well documented that [5,7-9], for PMN-PT compositions within the morphotropic phase boundary (MPB) range, the temperature $T_m$ of the permittivity maximum increases with $dc$ bias field, $E$, and the rate of this increase depends on the crystal orientation. Contrary to this, Viehland et al [14] showed that, in PMN-0.1PT ceramics, $T_m$ slightly decreases under small fields and strongly increases only at large fields. Similar observations have been reported for ceramic samples from 0.1≤$x$≤0.3 compositional range [21,22]. However, details of the $T_m(E)$ dependence for PMN-$x$PT crystals remain obscure.

In the present paper, we report experiments on the $dc$ bias field effect on the $T_m$ of [001] oriented PMN-$x$PT crystals including compositions with low titanium concentrations ($x$=0.06, 0.1 and 0.13) exhibiting relaxor behaviour both at zero bias and at bias field values, up to 5 kV/cm, as well as compositions adjacent ($x$=0.25 and 0.3) and belonging ($x$=0.35, 0.4) to the MPB compositional range, whose properties under the biased condition are similar to normal ferroelectric.

## II.   Experimental results

PMN-$x$PT single crystals used in this study were transparent plates cut from flux grown crystals prepared at the Physics Research Institute of the Rostov State University [3]. The large faces of the samples were perpendicular to [100] direction. Details of dielectric studies have been described elsewhere [4-6]. Zero field cooling (ZF), field cooling (FC), field heating after field cooling (FH), and zero field heating after field cooling (ZFHaFC) protocols were used.

Figure 1 shows the temperature dependences of the real ($\varepsilon'$) part of complex permittivity measured in the FC mode in the vicinity of the permittivity maximum for some of the crystals studied. Besides this main permittivity maximum (observed at $T=T_m$), the crystals with $x\geq 0.2$ exhibit at zero field additional anomalies at a lower temperature corresponding to the MPB or to a spontaneous phase transition from the relaxor to ferroelectric or mixed (ferroelectric/relaxor) state (see [5,6] for details). At the first glance, our results for the crystals with low ($x<0.2$) titanium content are in contradiction with the reports of the additional dielectric anomalies corresponding to the spontaneous transition from relaxor to ferroelectric phase in PMN-0.1PT ceramics [15] as well as with the appearance of the rhombohedral phase at low temperatures for $x\geq 0.05$ [16,17] as revealed by X-ray and neutron diffraction studies of PMN-PT powders. However, these contradictions seem to be due to surface effects. In single crystals and large-grain powders of PMN, the cubic phase is observed on cooling down to 10 K [10]. Besides, in fine-grained PMN powder, a rhombohedral macroscopic symmetry was detected [18]. Modern experimental technique, e.g. spatially resolved neutron diffraction, give some evidence that a very thin rhombohedral surface layer possibly exists in PMN, with a thickness much smaller than the penetration length of X-rays, so this skin could not be detected in the usual diffraction experiments in single crystals and coarse-grained ceramics [19,20].

In the crystals with low titanium content ($x=0.06$, 0.10 and 0.13, in the present study), at large enough bias, field-induced phase transitions from the relaxor to a ferroelectric or mixed state have been observed manifesting themselves in the FC mode by an abrupt drop (step) of dielectric permittivity. The temperature $T_S$ of this step depends on $E$. Below, we will mainly focus on this dependence as well as on the dependence of $T_m$ on $E$.

Fig. 2 shows the dependence of the reduced temperature $\Delta T_m=T_m(E)-T_m(E=0)$ on $E$. One can see that $T_m$ increases with $E$ in the high field regime, which is typical of usual ferroelectrics. For crystals with $x<0.4$, at low biases, in contrast to usual ferroelectrics, $T_m$ remains practically unchanged or even decreases with $E$ first, up to a certain threshold bias, $E_t$, above which $T_m$ increases. It is worth noting that such a type of the $T_m(E)$ dependence has been actually observed previously both in single crystals and ceramics of PMN-PT, as is evident from the analysis of the figures published, e.g. in Refs. [4,6,7,14,21,22]. As an example, panel "b" of Fig. 2 shows the $\Delta T_m(E)$ dependence for [111] oriented PMN-$x$PT crystals, constructed using the $T_m$ values deduced from the data published in Refs. [4,6,7]. The "anomalous" (from the point of view of ordinary ferroelectrics) low-field portion of this dependence has not been commented yet and was only scarcely mentioned in literature yet. The only exceptions are Refs. [21,23,24] where $E_t$ was considered as a field value necessary to overcome the action of random fields. This conception will be discussed in more details below.

Fig. 3 shows the concentration dependence of the field, $E_t$, at which $\Delta T_m(E)$ has a minimum position or starts abruptly increasing. Surprisingly, the $E_t(x)$ dependences for both [001] and [111] oriented crystals scale to one straight line. For comparison, we also plot the $E$ values corresponding to the inflexion points in the $\Delta T_m(E)$ curves for PMN-$x$PT ceramics, deduced from the data of Refs. [21,22]. One can see, that $E_t$ decreases with the increase of $x$, and the extrapolation of the linear part of $E_t(x)$ dependence both for ceramics and crystals intersects the $x$ axis at about $x$=0.33-0.35. However, some small, nearly constant, values of $E_t$ can be observed in the 0.3-0.35 compositional range. This remaining value seems to be due to compositional fluctuations typical of the PMN-$x$PT at compositions belonging to the MPB range [25].

Figs. 4 and 5 show the effect of $dc$ bias on the frequency dependence of the $T_m$ and $\varepsilon'(T)$ maximum magnitude ($\varepsilon_m$). In all the crystals studied, both the frequency dispersion of $\varepsilon_m$ and $T_m$ reduce drastically for biases exceeding $E_t$. This effect is largest for the PMN-$x$PT crystals in the MPB compositional range. This is consistent with numerous data showing that unpoled PMN-$x$PT crystals from the MPB compositional range exhibit relaxor-like behavior, while, at high fields, their properties are similar to ordinary ferroelectrics [6,7]. The new result is that there exists a threshold $dc$ bias value below which the behavior of the crystals is relaxor-like, while, above this threshold, it is more or less similar to usual ferroelectrics.

Fig. 6 shows the tentative $E$-$T$ phase diagrams for [001] PMN-$x$PT crystals with low titanium content. For comparison, the $E$-$T$ phase diagram for a [111] PMN crystal [4] is also presented and, in Fig. 7a, we show a common $E$-$T$ phase diagram for a system experiencing a first order phase transition. In Fig. 7a, the *FH* and *FC* lines correspond to the *appearance* of *macroscopic* metastable states and the $T_c$ line is the line where two phases (in our case, relaxor and ferroelectric) have equal free energy. Thermal hysteresis decreases with the increasing field and vanishes at the end point where the $T_c$, *FH* and *FC* lines intersect. Above this point, the ferroelectric phase is indistinguishable from the relaxor phase and dielectric permittivity has only a diffuse maximum, which depends little on frequency.

Vugmeister and Rabitz [12] modeled the $T_c$ line in the spirit of the Landau theory, however used the experimental temperature dependence of the quasistatic dielectric permittivity [26] obtained on cooling. They successfully fitted their semi-empirical expression to the relaxor-ferroelectric *T-E* boundary experimentally observed [4,5] on field cooling. Notice that a detailed analysis of the nucleation of the new phase at first order phase transitions [27] shows that a set of conditions should be satisfied in order the phase transition has happened at $T=T_c$ or close to this temperature. The validity of these conditions for PMN-*x*PT should be checked in future studies.

It is seen from the plotted experimental phase diagrams that, in the FC mode, for PMN-*x*PT crystals (*x*>0), there appears a nearly vertical portion, above the relaxor-ferroelectric boundary, which has not been observed in pure PMN. This portion looks like an additional vertical line in the *T-E* phase diagram and, in order to check this possibility, we consider additional scalar order parameter, *Q*, which is coupled to the polarization, *P*, with a positive coupling constant (c.f. [3]) and with other constants assuming a first order phase transition under bias field (Fig. 7b). The appearance of *Q*, in this model, triggers the gain of polarization and, due to coupling between *Q* and polarization, the phase transition is first order. The experimentally observed vertical portion in the phase diagram looks similar to the vertical line in Fig. 7b although we have not found experimental evidence of the phase transition below the relaxor-ferroelectric boundary. It is possible that this is because random fields smear this phase transition at small bias fields.

It is important that coupling between *P* and *Q* changes the FC line and, in particular, there appears a threshold field [28], below which there are no phase transitions in agreement with experiment.

The meaning of the order parameter *Q* is not clear, for the moment. We assume a close relationship to the dynamic behavior of so-called "polar nanoregions" (PNRs), which are believed to be at the heart of the relaxor behavior [1,29]. We should notice that this line is close to the Vogel-Fulcher temperature, $T_{VF}$, where the PNR's size strongly changes [30] and above which the PNRs are dynamic ($T_{VF}$ is always close to the ZFHaFC line, which marks the appearance of macroscopic metastable states). At low bias fields, the phase transition is diffuse due to coupling between *Q* and random fields [13]. Thus, *Q* might refer to the average size of PNR [30], the Edwards-Andersson parameter [31] or any other scalar order parameter with the above described properties.

Such a quasi-vertical line has been observed also in ZFHaFC experiments in PLZT ceramics (see [32] and references therein). This line was connected with depolarization resulting in the spontaneous weak first order phase transition from the ferroelectric to the ergodic relaxor phase. One can explain this by PNR's defreezing and by vanishing of the metastable macroscopic state in free energy at the ZFHaFC boundary responsible for the nucleation of the ferroelectric phase.

## III. Discussion

A widely accepted complete description of relaxors [28] considers dipoles under the action of random bonds and in random fields, which were introduced into the discussion of relaxor behavior previously [33,34]. In the present study, we concentrate on the bias field dependence of the diffuse peak. Below, we will test different models considering nonlinear dielectric permittivity in PMN, in connection with our experimental results.

Vugmeister and Rabitz described the frequency and bias field dependence of FC dielectric susceptibility [12,35] in relaxors by considering the average polarization produced by PNRs in the spirit of the Landau mean-field theory. The main expression can be also obtained from the theory of nonlinear susceptibility of coupled dipoles [36]:

$$\varepsilon = \varepsilon_\infty + \frac{\chi_s}{1 - \lambda^2 n \chi_0 F} \qquad (1)$$

where

$$\chi_0 = \left\langle \frac{1}{\alpha + 3\beta P^2 + 5\gamma P^4 + 7g P^6} \right\rangle \qquad (2)$$

$$F = \frac{1}{4 k_B T \cosh^2(u / k_B T)} \left\langle \frac{1}{1 + i \omega \tau} \right\rangle \qquad (3)$$

The polarization $P$ can be found from the equilibrium condition:

$$\alpha P + \beta P^3 + \gamma P^5 + g P^7 = E + e \qquad (4)$$

where $e$ is random field given by one of the following two distribution functions:

$$g(e) = \frac{1}{2} \left[ \delta(e - e_0) + \delta(e + e_0) \right] \qquad (5)$$

$$g(e) = \frac{1}{\sqrt{\pi}} e^{-a e^2} \qquad (6)$$

The angular brackets in expression (2) mean averaging over the random field magnitude while the angular brackets in expression (3) imply averaging over the barriers.

In expressions (1-3), $u = 2\mu E + \lambda P$, $\mu$ is the dipole moment, $\lambda$ is a constant coupling the dipole moment to the polarization $P$, $n$ is the concentration of dipoles, $\chi_s$ is the low-frequency susceptibility without coupling to the average polarization, $\varepsilon_\infty$ is the high-frequency dielectric permittivity, $\tau = \tau_0 \exp\left[U / k_B \left(T - T_{VF}\right)\right]$. The Landau coefficient $\alpha$ is assumed to depend on $T$ in the manner, which has been described by Potts-type model computations of PMN [37] as well as by the SRBRF model and experiment [26,28]: it decreases linearly with decreasing temperature and saturates at low temperatures without crossing zero. Correspondingly, due to random fields and random bonds, the correlation length increases as temperature decreases and, then, saturates at low temperatures.

Our analysis has shown that the main factor influencing the direction of the shift of $T_m$ under bias field is the direction of the shift of susceptibility (2). If $\beta > 0$, like as in ferroelectrics with a second order phase transition, and random fields were absent then the maximum of $\chi_0(T)$ would shift upward, in small fields. This would contradict experiment. If $\beta < 0$ then this maximum shifts downward, first, and upward, above some threshold field. So, one can see that the result seems to depend crucially on the sign of $\beta$.

Experiments [38] show that $\beta$ is strongly anisotropic with respect to the main axis. For PMN at about 250 K, $\beta$ is small for the [111] oriented crystals, but it is comparatively large and positive for [001] orientation. Indeed, our experiment exhibits a dip for [111] oriented PMN but, for the most of the [001] oriented PMN-$x$PT, $T_m$ is kind of insensitive to the field below the threshold. At the first glance, it seems that the positive sign of $\beta$ contradicts the experimentally observed first order phase transition between the relaxor and ferroelectric phase. Pirc, Blinc, and Kutnjak [13] assumed that the parameter, which is responsible for the average interaction among the dipoles, $J_0$, increases with the increasing bias field, and just this makes the relaxor-ferroelectric phase transition possible at finite fields, in PMN. Vugmeister and Rabitz [12] used in their model a positive value of $\beta$, but $\gamma$ was negative. We computed the dependence $\chi_0(E)$ without averaging over random fields at different reasonable values of $\beta$ and found that, $\beta$ does not play a significant role at $T = T_m$, in PMN (because $3\beta P^2$ is comparatively small with respect to $\alpha$ at actual fields), and $\chi_0$ is approximately constant until about 3 kV/cm. This fact would corresponds to silence of $T_m$ to bias field in this range. Above 3 kV/cm, $T_m$ becomes strongly inclined in the direction of the end point in the phase diagram if one supposes a first order phase transition. Above the end point, the dependence of $T_m(E)$ changes again, and $T_m(E)$ starts increasing. At comparatively high fields, this dependence is consistent with our experiment but, at low $E$, the experiment performed for PMN shows the existence of a dip that is not fully reproduced by the considered theory. We found that taking into account random field distribution (5) reproduces this dip (Fig. 8a), while distribution (6) provides the dependence, which is insensitive to $E$ (Fig. 8b). Similar effect of random fields on $T_m(E)$ at second order phase transitions was discovered earlier by Dorogovtsev [23] (c.f. the result of averaging of polarization with the same distribution function [39]).

In the first case [distribution (5)], the quenched fields are up and down, and the main contribution to the shift of $T_m$ stems from the random fields, which are opposite to external field. $T_m$ has a dip in this case until the field reaches the magnitude of the random field, $e = e_0$, and, then, $T_m$ increases as it happens in normal ferroelectrics with a second order phase transition ($\beta > 0$, $\gamma > 0$). In the case if $\gamma < 0$ (see [12]), the dip continues until coming close to the end point in the phase diagram (the lower the frequency the closer is $T_m$ to the end point), and only then $T_m$ starts increasing. This behavior suits qualitatively our experiment and explains not only the existence of the dip in some cases but also the fact that $T_m(E)$ is directed towards the end point for all considered concentrations. Thus, our results are in favor of distribution (5), at least for pure PMN with [111] orientation. Also, Monte-Carlo computations performed for a 2D set of dipoles with random fields and random bonds gave a shallow dip in $T_m(E)$ until a threshold field, above which $T_m$ started increasing [24]. These data emphasize the necessity of using the random field – random bond idea in order to explain the low-field dependence of $T_m$ in PMN-$x$PT.

Now, we want to discuss briefly other possible models. Poplavko [40] assumed that the average relaxation time in relaxors decreases with the increase of bias field. Tagantsev and Glazounov [39] suggested that the vibration of PNRs can be described as $\tau = \tau_0 \exp\left[(U - VPE)/k_B T\right]$ where $V$ is the PNR's volume and $P$ polarization magnitude inside PNR's. These suggestions lead to decrease of the relaxation time with $E$. Though they would be in line with our finding, experiments performed on ceramics [14] do not confirm the decrease of the relaxation time in PMN below the threshold field. As another example, in the SRBRF model [13], the frequency dependent part of dielectric permittivity is expressed over $z + i\omega\tau$ where $z$ is a model parameter, which can be found as a function of the average dipole-dipole interaction energy, $J_0$, and average dispersion of this interaction, $J$. When $z$ increases, the relaxation time decreases, and $T_m(E)$ moves downward. Our analysis has shown that $z$ only slightly increases at small fields but this increase becomes rather strong at the

relaxor – ferroelectric boundary. Thus, this model predicts that the relaxation time only slightly changes at small fields that is consistent with experiment [14].

## IV.   Summary

Our experiment has shown that, in all studied PMN-$x$PT single crystal compositions, there is a FC line in the $E$-$T$ phase diagram separating the relaxor and ferroelectric phases. A first-order phase transition happens at this line. In contrast to pure PMN, this line has a portion, which does not depend on the bias field (quasivertical line), which we attribute to a first-order nonpolar phase transition. In PLZT, the same line has been observed on ZFHaFC and it was shown that depolarization takes place at this line [32,41].

The temperature of the maximum of the dielectric permittivity, $T_m$, was found to be silent to the bias field or moved downward at small fields and increased at large fields. Our analysis of different models of relaxors has shown that, in order to explain this effect, a first order phase transition at the relaxor – ferroelectric phase boundary has to be taken into account as well as quenched random fields and random bonds.  The quenched uncorrelated fields, which are at the origin of the generic random-field model [33] relate in relaxors to their inherent charge disorder [34]. The Dorogovtsev effect extended to the relaxor systems provides a good qualitative description of the dependence of $T_m(E)$ at small fields. At large fields, both the Vugmeister and Rabitz model [12] supplemented with a temperature dependence of quasistatic dielectric permittivity and the Pirc, Blinc, and Kutnjak model [13] explaining both the quasistatic dielectric permittivity, frequency, and the bias field dependences provide a reasonable explanation of our experiment.

This study is partially supported by RFBR (grants 04-02-16103 and 05-03-32214). I. P. R. appreciates a financial support from a Scientific Research Fellowship and hospitality of LPMC, Universite de Picardie Jules Verne. I.P.R. and S.A.P. are grateful to V.Yu.Topolov and V.P. Sakhnenko for discussions.

## Figure Captions

Fig.1 $\varepsilon'(T)$ dependences measured in the FC mode in the vicinity of the permittivity maximum for some flux-grown (001) PMN-*x*PT crystals at different *E* values (in each panel *E* grows from the leftmost curve to the rightmost one).

    a)     *x*=0.06; *f*=1 kHz; *E*=0;1;2;4 kV/cm;

    b)     *x*=0.13;*f*=1 kHz; *E*=0;0.25;0.5;1;2;3.5;4 kV/cm;

    c)     *x*=0.35; *f*=10 kHz; *E*=0;0.1;0.4;1;1.5;2;2.5;3 kV/cm;

    d)     *x*=0.4; *f*=1kHz; *E*= 0;0.5;1;1.5; 2;2.5;3 kV/cm;

Fig. 2 Dependences of the reduced temperature $\Delta T_m = T_m(E) - T_m(E=0)$ on *dc* bias *E* for PMN-*x*PT crystals. The numbers correspond to *x* values.

a) for [001] crystals studied in the present work.

b) for [111] crystals, constructed using the $T_m$ values deduced from the data published in Ref. [4,8,9].

Fig. 3. The concentration dependence of the field, $E_t$, at which $\Delta T_m(E)$ has a minimum position or starts abrupt increasing, for [001] (triangles) and [111] (circles) PMN-$x$PT crystals and for ceramics (squares), estimated from the minimum (empty symbols) or inflection (filled symbols) in the $T_m(E)$ curves. The data for [111] crystals are deduced from Ref. [4,6,7] and for ceramics-from Refs. [21,22].

Fig.4. $T_m(E)$ dependences for PMN-(0.13)PT crystal measured at different frequencies: 1; 5; 10; 50;100 kHz (from bottom to the top).

Fig.5. The effect of bias field on the frequency dependences of $T_m$ (a) and $\varepsilon'_m$ (b) for the PMN-$x$PT crystals: $x$=0.06 (1); 0.13 (2); 0.25 (3); 0.35 (4).

$\Delta T_m = T_m(100 \text{ kHz})- T_m(1 \text{ kHz});$

$\Delta\varepsilon'_m /\varepsilon'_m = [\varepsilon'_m(1 \text{ kHz})-\varepsilon'_m(100 \text{ kHz})]/\varepsilon'_m(1 \text{ kHz}).$

Fig. 6. $E$-$T$ phase diagrams for some PMN-$x$PT crystals: a) [111] PMN [4], b) $x$=0.06, c) $x$=0.10, d) $x$=0.13. The dashed and dotted lines are just guides for the eye and are mostly hypothetical.

Fig. 7. Fig. 7. Theoretical modelling: (a) The phase diagram for a system experiencing a first-order phase transition described within a Landau theory; (b) a phase diagram including ferroelectric ($P$), and Q phases where $Q$ is a scalar order parameter coupled to polarization $P$.

Fig. 8. The dependence of $T_m(E)$ obtained within a model, which takes into account random fields. Panels a and b correspond to different distribution functions of random fields (see text).

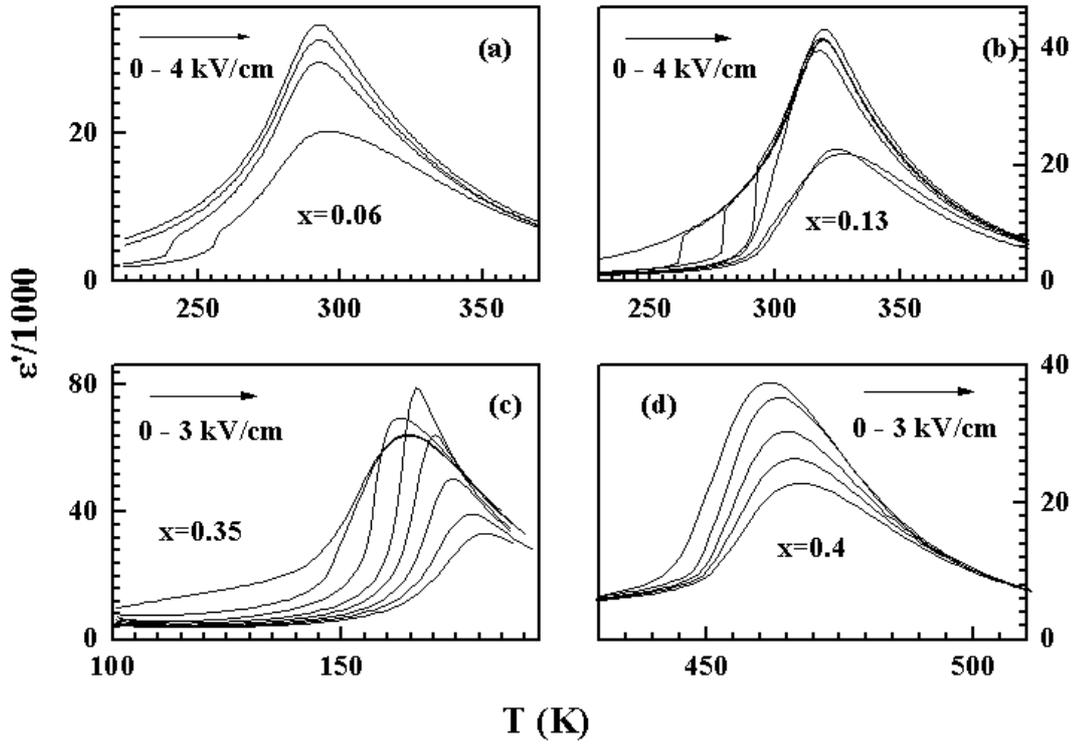

Figure 1

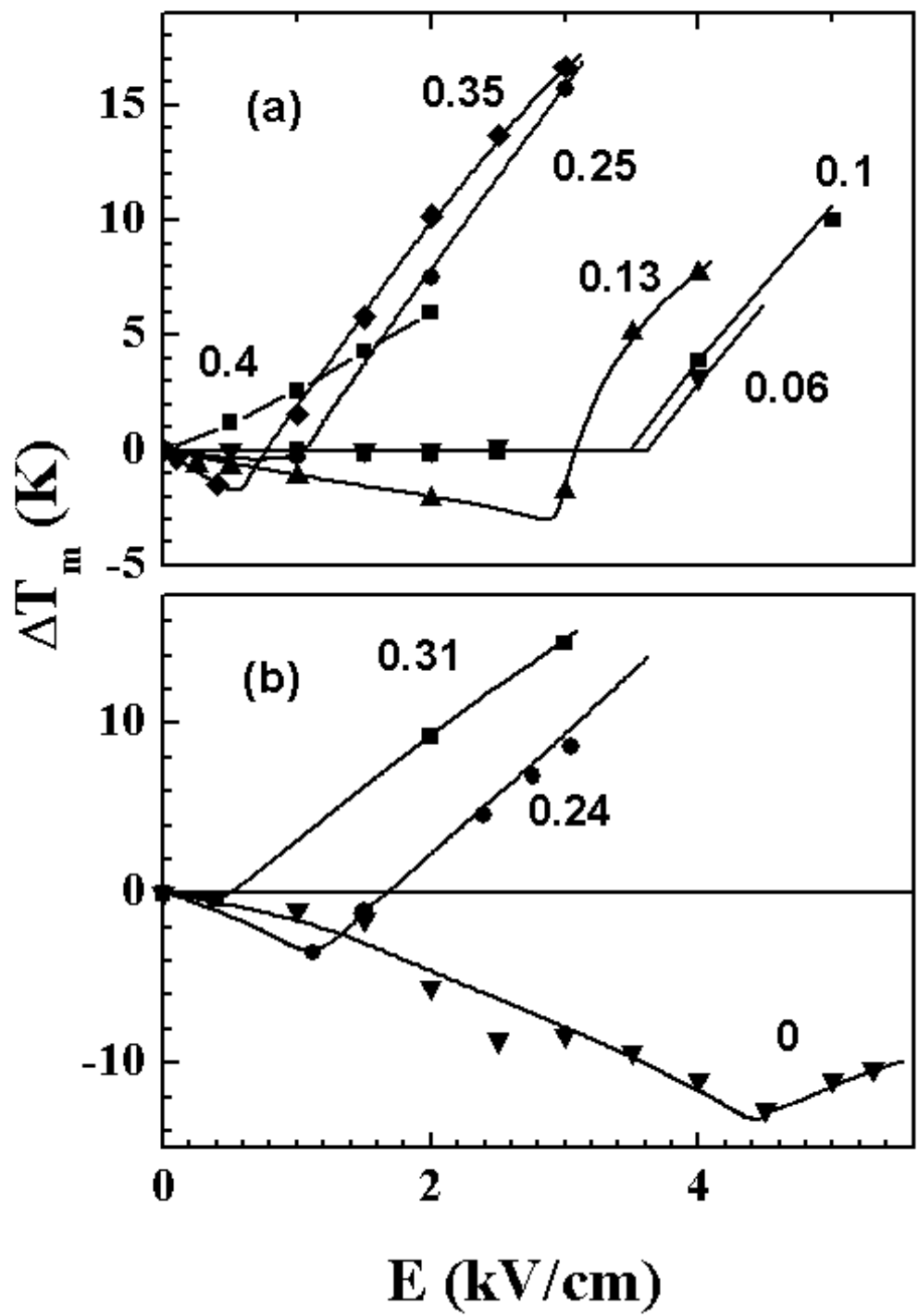

Figure 2

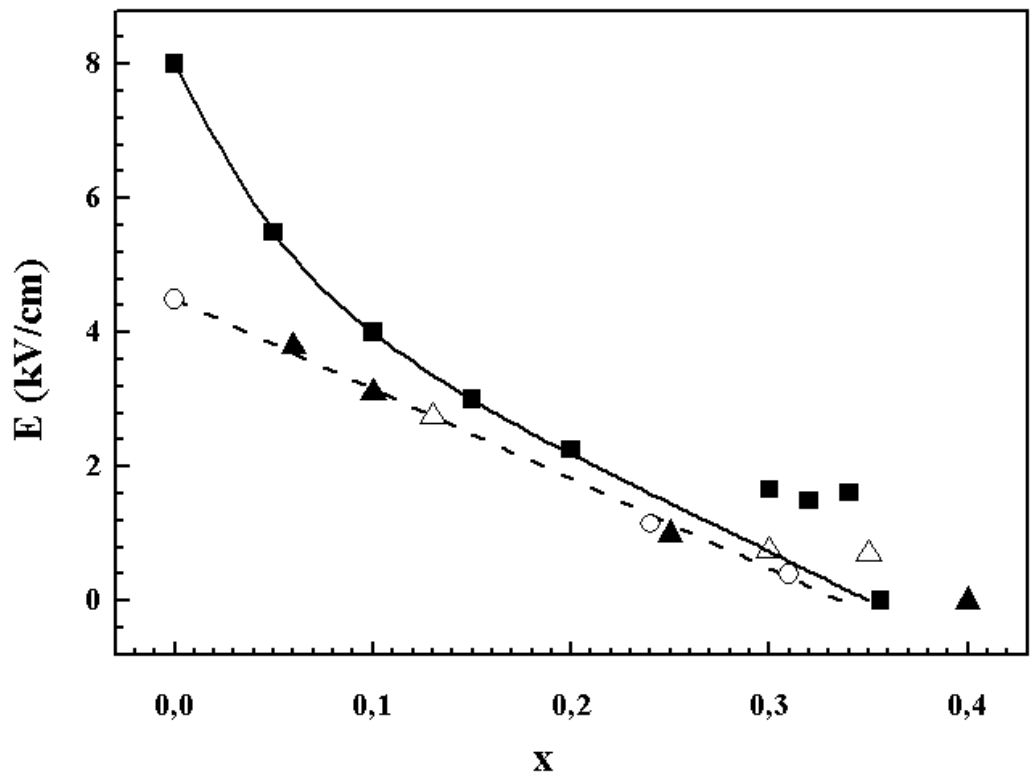

Figure 3

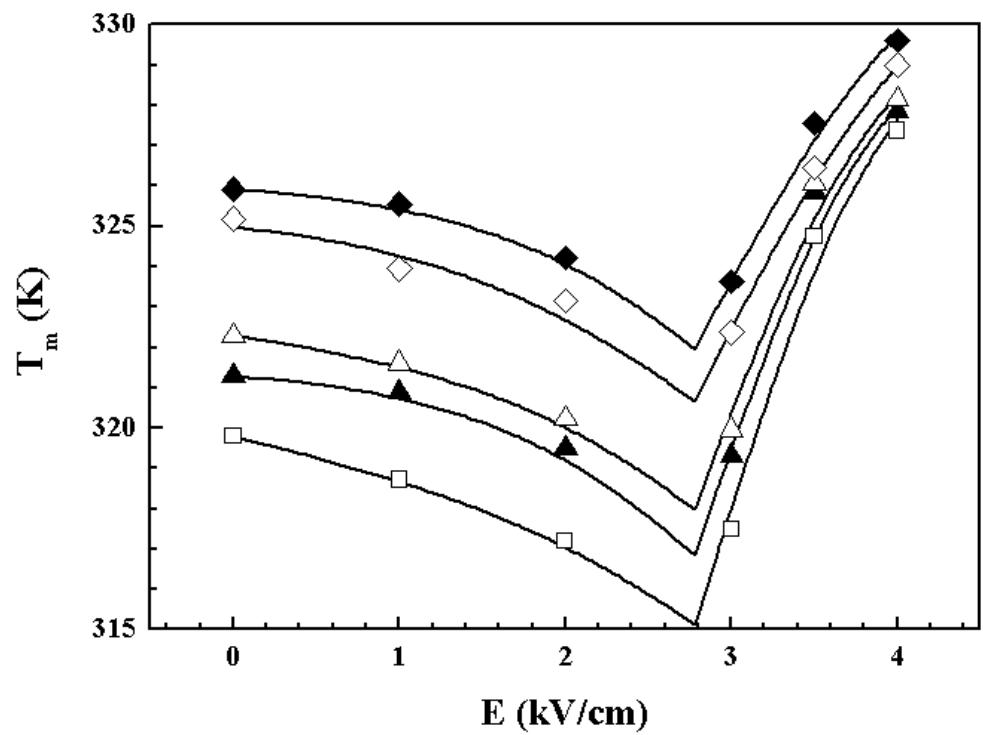

Figure 4

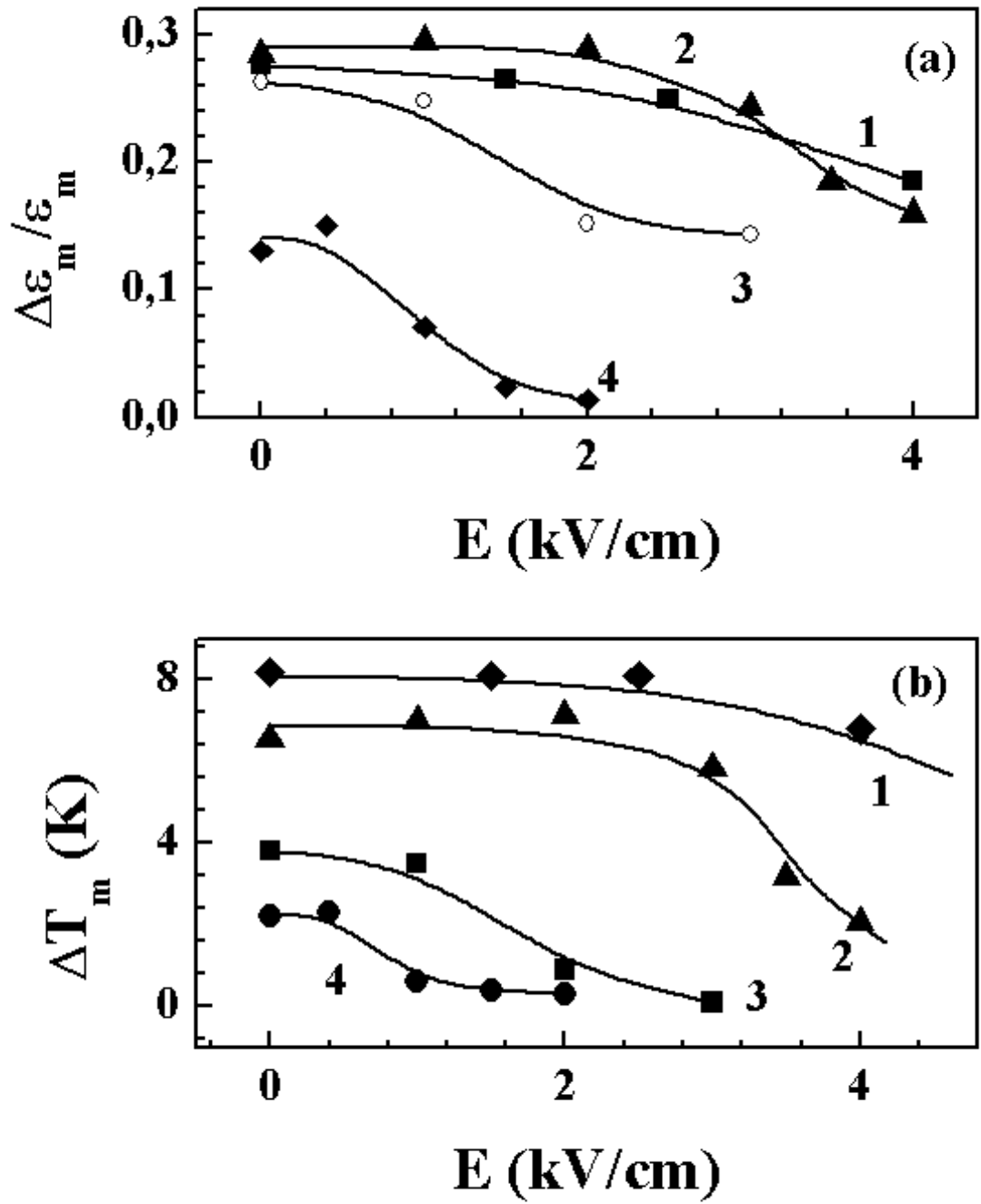

Figure 5

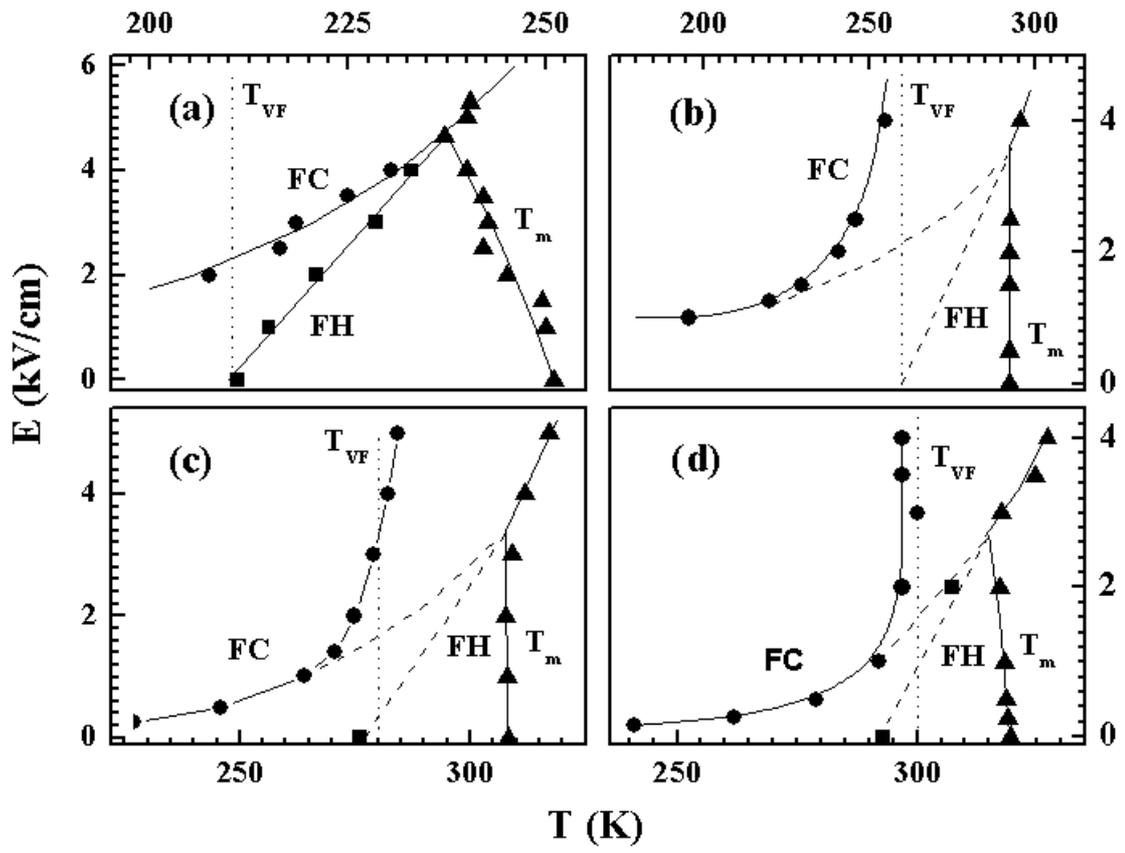

Figure 6

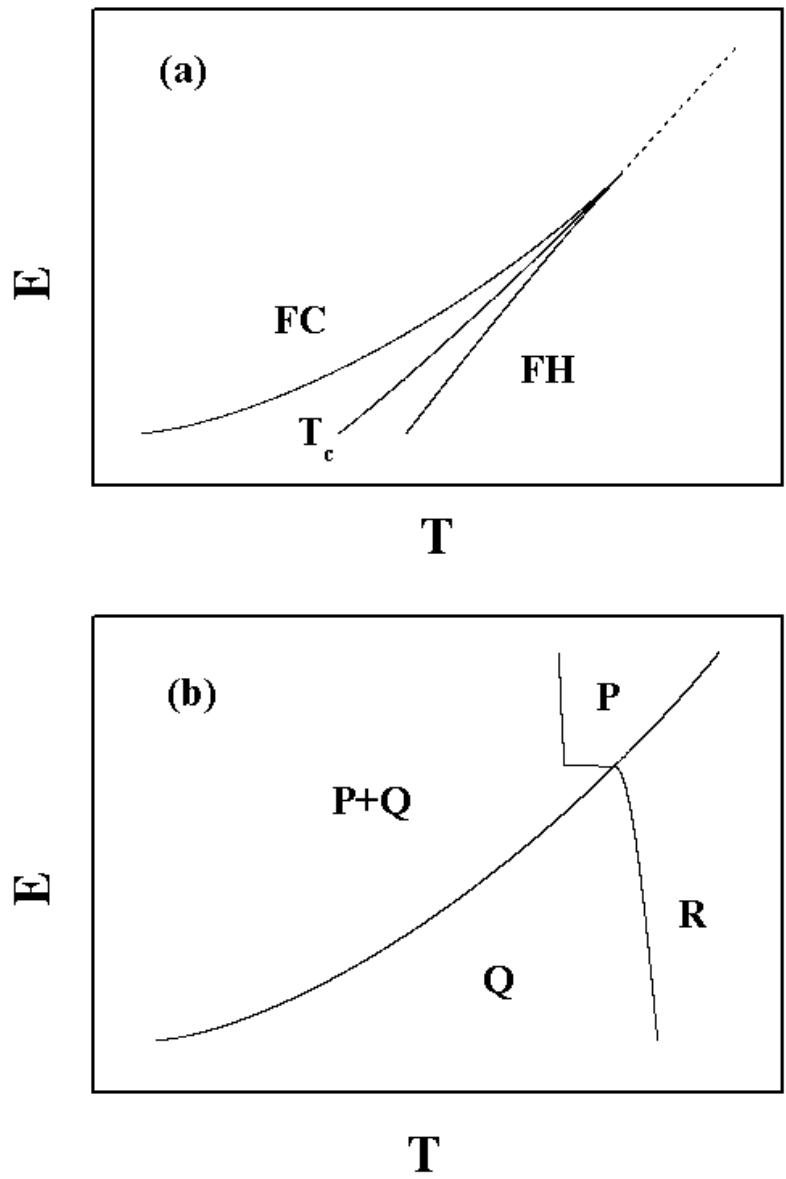

Figure 7

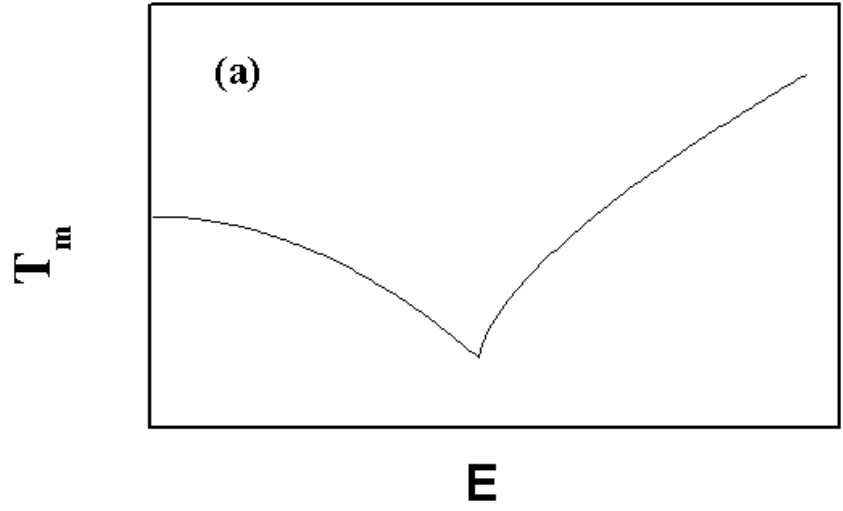

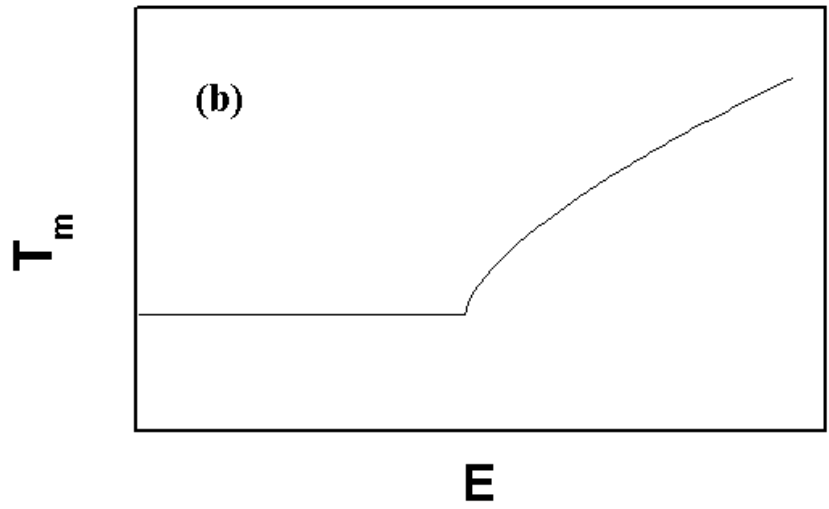

Figure 8